\documentclass{aa}
\usepackage{graphicx}
\usepackage{natbib}
\usepackage{txfonts}
\bibpunct{(}{)}{;}{a}{}{,}

\begin{document}

\hyphenation{be-san-con mo-dele }

\title{Early emission of rising optical afterglows:
\\
The case of \object{GRB~060904B} and \object{GRB~070420}.
\thanks{Based on observations performed with
TAROT at the Calern/OCA and La Silla/ESO observatories,
GCN data archive and \emph{Swift} public data archive.
}
}

\author{A. Klotz\inst{1,2}
        \and B. Gendre\inst{3,4}
        \and G. Stratta\inst{5}
        \and A. Galli\inst{3,6}
        \and A. Corsi\inst{3}
        \and B. Preger\inst{5}
        \and S. Cutini\inst{5,7}
        \and A. P\'elangeon\inst{8}
        \and J.L. Atteia\inst{8}
        \and M. Bo\"er\inst{1}
        \and L. Piro\inst{3}
}

\institute{
Observatoire de Haute-Provence, F--04870 Saint Michel l'Observatoire, France
\and CESR, Observatoire Midi-Pyr\'en\'ees, CNRS, Universit\'e de Toulouse, BP 4346, F--31028 - Toulouse Cedex 04, France
\and Istituto di Astrofisica Spaziale e Fisica Cosmica, Sede di Roma, INAF, Via Fosso del Cavaliere 100, I--00133 Roma, Italy
\and University degli Studi di Milano - Bicocca - Piazza dell'Ateneo Nuovo, I--20126, Milano
\and ASI Science Data Center (ASDC), via G. Galilei, I--00044 Frascati, Italy
\thanks{
INAF personnel resident at ASDC under ASI contract 1/024/05/0.
}
\and INFN-Trieste, Padriciano 99, I--34012 Trieste, Italy
\and Universit\`a di Perugia, Dipartimento di fisica, Viale A. Pascoli, I--06123 Perugia, Italy
\and LATT, Observatoire Midi-Pyr\'en\'ees, CNRS, Universit\'e de Toulouse, 14 Avenue E. Belin, F--31400 - Toulouse, France
}

\offprints{A. Klotz, \email{klotz@cesr.fr}}

\date{Received {\today} /Accepted }

\titlerunning{Early emission of \object{GRB~060904B} and \object{GRB~070420}}
\authorrunning{Klotz {\it et al.}}

\abstract
   {}
   {We present the time-resolved optical emission of gamma-ray bursts \object{GRB~060904B} and \object{GRB~070420} during
   their prompt and early afterglow phases.}
   {We used time resolved photometry from optical data taken by the TAROT telescope
   and time resolved spectroscopy at high energies from the \emph{Swift} spacecraft instrument.}
   {The optical emissions of both GRBs are found to increase from the end of the prompt phase,
   passing to a maximum of brightness at $t_{peak}$=9.2\,min and 3.3\,min for
   \object{GRB~060904B} and \object{GRB~070420} respectively and then decrease.
   \object{GRB~060904B} presents a large optical plateau and a very large X-ray flare.
   We argue that the very large X-flare occurring near $t_{peak}$ is produced
   by an extended internal engine activity and is only
   a coincidence with the optical emission. \object{GRB~070420} observations would support
   this idea because there was no X-flare during the optical peak. The nature
   of the optical plateau of \object{GRB~060904B} is less clear and might be related to the
   late energy injection.
   }
   {}


\maketitle

%%%%%%%%%%%%%%%%%%%%%%%%%%%%%%% INTRODUCTION %%%%%%%%%%%%%%%%%%%%%%%%%
\section{Introduction}

Since the discovery of Gamma-Ray Burst (GRB) X-ray afterglows, in 1997 \citep{costa97},
it has been possible to obtain a precise burst localization that allows the
follow-up of GRBs from radio to X-ray \citep{wijers99}.
From that date, tens of GRB optical afterglows have been detected
by ground-based rapid response telescopes.
The launch of the \emph{Swift} satellite, in late 2004 \citep{Gehrels2004}
has increased the usefulness of robotic telescopes.
In fact, the capability of \emph{Swift} to alert other observatories
within seconds after the burst, and then re-point quickly on the burst position
has allowed us to significantly increase the number of successful afterglow
optical detections and to perform early optical observations.
The large amount of data collected during these years has shown
that during the first phases (a few hundreds of seconds) the
optical emission can be as bright as $m_V \sim 14-17$ before
decreasing as $F(t) \propto t ^{-\alpha} \nu ^{-\beta}$,
with ${\alpha}{\sim}1.1-1.7$ \citep{fox05,berger05,depasquale06,Gendre2006}.
Several events were  observed optically during the prompt
phase\footnote{We call 'prompt phase' the
period during which \emph{Swift}-BAT detected the high energy emission defined by the $T_{90}$ duration.},
{\it e.g.} \object{GRB~990123} \citep{akerlof99},
\object{GRB~041219A} \citep{Vestrand2006},
\object{GRB~050401} \citep{rykoff05},
\object{GRB~050904} \citep{boer06}
or \object{GRB~060111B} \citep{Klotz2006b}.
In some cases the optical and $\gamma$-ray prompt emissions appeared
 correlated ({\it e.g.} \object{GRB~041219A}) and are likely produced by
internal shocks generated by variations in the energy ejection due
to the central engine \citep{Vestrand2006}.
In other cases ({\it e.g.} \object{GRB~990123}) the optical prompt emission is highly
variable and not correlated with the $\gamma$-ray one; they can be
produced by different source regions ({\it i.e} internal and reverse
shocks respectively), and/or by different emission mechanisms
(Inverse Compton and synchrotron) \citep{akerlof99}.

Early optical afterglow data play a role in obtaining
information on the physics of the central engine,
and possibly in constraining the initial Lorentz factor
of the fireball \citep[{\it e.g} ][]{zhang03}.
However, the beginning of
the optical afterglow emission is often missed either because it occurs
before the first observations are completed, or because it is hidden by the
high brightness of an optical flash that sometimes occurs when
the GRB is still active \citep[{\it e.g}
\object{GRB~060111B} in][]{Klotz2006b}.
Sometimes, the optical afterglow takes a few minutes to reach its maximum
and rapid response optical telescopes are able to record the rise
of the light curve. For example this is the case of \object{GRB~060418}
and \object{GRB~060607A} \citep{molinari2007},
which allowed investigators to constrain the initial
fireball Lorentz factor to $\Gamma_0 \sim 400$.

In this paper we report the early optical observations
from 23~sec to 43 min of \object{GRB~060904B} and
from 40~sec to 18 min of \object{GRB~070420} performed with the TAROT
robotic observatories.
These two GRBs have an optical afterglow
that reaches a maximum hundreds of seconds after the GRB.
We compare early simultaneous optical, X-ray and $\gamma$-ray data
to address the question of the relation between optical and high
energy emissions in the first minute after the GRB.

\subsection{GRB060904B}

\object{GRB~060904B} was detected on September 4$^{th}$, 2006
at 02:31:03 UT
by the BAT instrument on the \emph{Swift} spacecraft
\citep[trigger=228006,][]{Grupe2006}.
This GRB is a double--peak event \citep{Markwardt2006}:
the BAT light curve shows an initial
fast rise exponential decay pulse 9 seconds wide
at 2 seconds before the BAT trigger
(hereafter t$_\mathrm{trig}$)
followed by a second, weaker peak detected within the 15-25 keV range, starting at
t$_\mathrm{trig}$+120 sec, peaking at t$_\mathrm{trig}$+155 sec and finishing
at t$_\mathrm{trig}$+220 sec. $T_{90}$ (15-350 keV) is 192$\pm$5 sec.
\emph{Swift}-XRT observations began at t$_\mathrm{trig}$+69 sec. The X-ray light curve
displays a shallow emission during the gamma emission, followed by a huge flare.
Optical ground follow-up began very early when the gamma emission
was still active:
ROTSE-IIIc \citep{Rykoff2006} detected a source
at RA=03$^h$52$^m$50.52$^s$ Dec=-00$^\circ$43'30.85" (J2000,0)
with R=17.3 at t$_\mathrm{trig}$+18.5 sec.
\citet{Klotz2006a}
reported a brightening of the optical transient reaching
a peak of brightness R$\sim$17 near t$_\mathrm{trig}$+400\,sec.
Optical spectroscopy was performed with VLT + FORS1 \citep{Fugazza2006},
at t$_\mathrm{trig}$+5.15 hours observing several metallic absorption lines at $z$=0.703.
\citet{Grupe2006} mentioned that
\emph{Swift}-UVOT detected the afterglow in a finding chart exposure of 246 seconds
with the V filter started 70 seconds after the BAT trigger.
According to \citet{Schlegel1998} galactic extinction is E(B-V)=0.172
thus implying (assuming $R_V$=3.1) $A_B$=0.74, $A_V$=0.57 and $A_R$=0.46.

\subsection{GRB070420}

\object{GRB~070420} was detected on April 20$^{th}$, 2007
at 06:18:13 UT by the \emph{Swift}-BAT instrument
\citep[trigger=276321,][]{Stamatikos2007}.
\emph{Swift} instrument analysis was reported by \citet{Stamatikos2007b}
providing celestial coordinates from UVOT at
RA=08$^h$04$^m$55.17$^s$ Dec=-45$^\circ$33'20" (J2000,0).
The BAT light curve shows a slow rise that began 50\,sec before
the trigger.
$T_{90}$ (15-350 keV) is 77$\pm$4 sec.
\emph{Swift}-XRT observations began at t$_\mathrm{trig}$+99 sec. The X-ray light curve
displays a steep decay
between t$_\mathrm{trig}$+106 and t$_\mathrm{trig}$+300 sec. Then, a plateau
occurred until at least t$_\mathrm{trig}$+2000 sec before a decay until
at least few 10$^4$ sec. No X-flare was detected.
\citet{DAvanzo2007} observed the transient at
H=13.1$\pm$0.2 with the REM telescope.
From high energy spectral properties delivered by the KONUS
experiment,
\citet{Pelangeon2007} computed a pseudo-redshift $pz$=1.56$\pm$0.35.
UVOT detection in U the band \citep{Immler2007} confirms the low redshift of the burst.
According to \citet{Schlegel1998} galactic extinction is E(B-V)=0.50.
Assuming $R_V$=3.1, the extinctions are thus
$A_B$=2.18, $A_V$=1.63, $A_R$=1.33, $A_H$=0.29.

%%%%%%%%%%%%%%%%%%% FIG image %%%%%%%%%%%%%%%%%%%
\begin{figure}[htb]
\centering{\includegraphics[width=0.9\columnwidth]{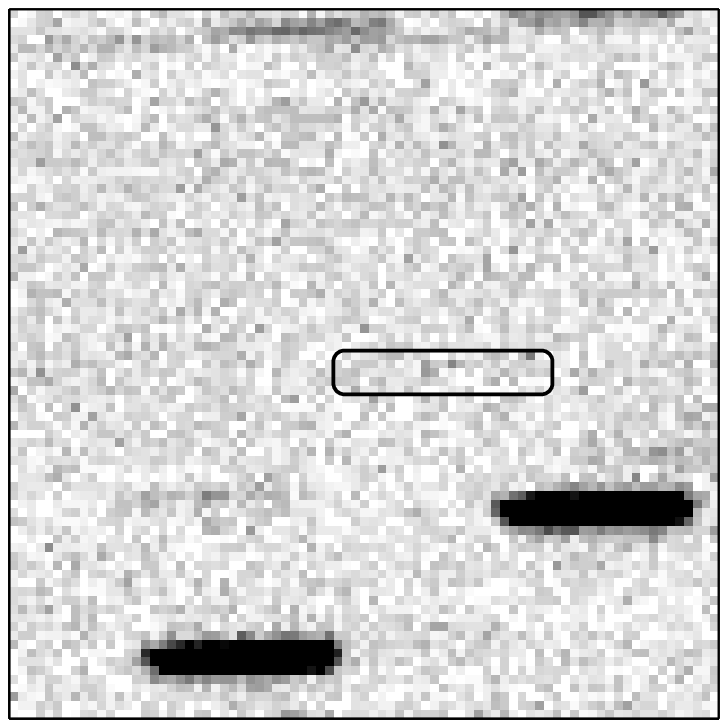}}
\centering{\includegraphics[width=0.9\columnwidth]{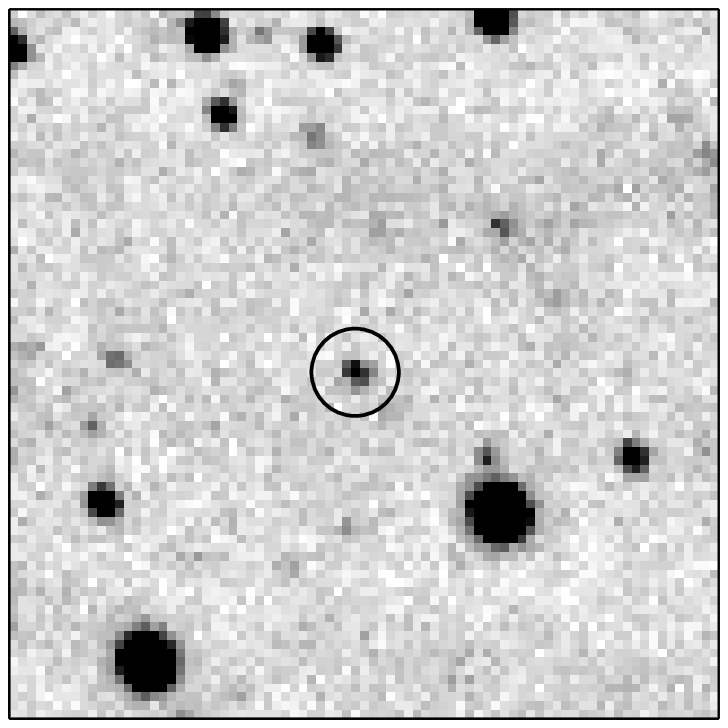}}
\caption{
\object{GRB~060904B}. Top: This image was taken between 23\,sec and 83\,sec after the GRB trigger.
The hour angle velocity was adapted to obtain stars as trails
of $\sim$20 pixel length during the 60\,sec exposure.
The theoretical position of the GRB trail is indicated by the rectangle.
A cluster of 4 pixels is present near the center of the rectangle
but is considered as noisy pixels or "cosmic" trace rather than an optical flash
(see text).
Bottom: Sum of images taken by TAROT in diurnal motion mode. The optical transient
position is indicated by the circle. The image size is 7 arcmin,
North is up, East left.
} \label{trail1}
\end{figure}
%%%%%%%%%%%%%%%%%%%%%%%%%%%%%%%%%%%%%%%%%%%%%%%%%%%%%%%%

%%%%%%%%%%%%%%%%%%% FIG image %%%%%%%%%%%%%%%%%%%
\begin{figure}[htb]
\centering{\includegraphics[width=0.9\columnwidth]{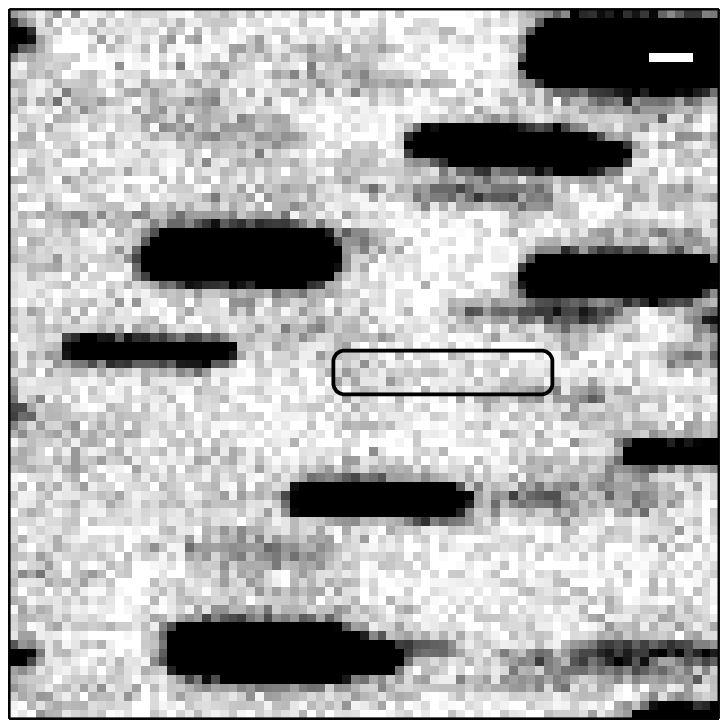}}
\centering{\includegraphics[width=0.9\columnwidth]{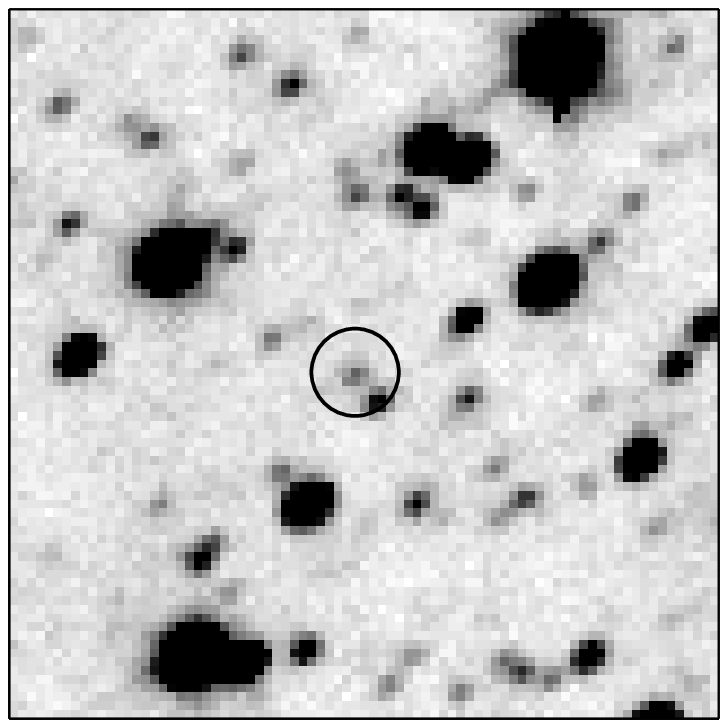}}
\caption{
\object{GRB~070420}. Top : This image was taken between 34\,sec and 94\,sec after the GRB trigger
in drift mode (see Fig.~\ref{trail1}). Three nearby stars
(see section~\ref{imageprocessing}) have been
subtracted.
The position of the GRB trail is indicated by the rectangle.
A group of faint lightened pixels is present in the first third of the rectangle
but we cannot conclude on its significance.
Bottom: Sum of images taken by TAROT in diurnal motion mode.
The optical transient
position is indicated by the circle (nearby stars are not
subtracted). The image size is 7 arcmin,
North is up, East left.
} \label{trail2}
\end{figure}
%%%%%%%%%%%%%%%%%%%%%%%%%%%%%%%%%%%%%%%%%%%%%%%%%%%%%%%%

%%%%%%%%%%%%%%%%%%%%%%%%%%%%%%% OBSERVATIONS %%%%%%%%%%%%%%%%%%%%%%%%%
\section{TAROT observations}
\label{obsdata}

TAROT are two fully autonomous 25 cm aperture
telescopes installed at Calern observatory (Observatoire de la
C\^ote d'Azur - France) and at La Silla observatory (European
Southern Observatory - Chile). These telescopes are
devoted to very early observations of GRB~optical counterparts.
A technical description can be found in \citet{Bringer2001}.
\object{GRB~060904B} was observed with TAROT Calern and \object{GRB~070420}
with TAROT La Silla.
\\
When the notice is received, the first image
is an unfiltered 60\,sec exposure taken in drift mode
\citep[see Figs.~\ref{trail1} and~\ref{trail2} top,
described in][]{Klotz2006b}.
For each GRB, the image began during the prompt
phase with a drift velocity of 3\,sec/pixel.
Successive images were tracked on the diurnal
motion. A first series of five unfiltered
(hereafter filter C) 30\,sec images was followed by four series of three 90\,sec images
filtered C--C--R respectively. Finally, series of three 180\,sec images were
taken with the same filter alternance.
In the \object{GRB~060904B} case, the drift image
began at t$_\mathrm{trig}$+23\,sec and entirely spans
the prompt phase (that extends until 220 sec after the trigger).
Dawn arrived 43 minutes after the trigger and observations were stopped.
For \object{GRB~070420}, the drift image
began at t$_\mathrm{trig}$+34\,sec. The
field elevation decreased from 8 degrees above horizon
to 5 degrees at the end of observations.
Only the first 18~minutes of the event were recorded.
Logs of observations are included in 
Tables~\ref{logobstable1} and~\ref{logobstable2}.

\subsection{Image processing}
\label{imageprocessing}

Photometry was performed using Point Spread Function (PSF) 
fitting taking a nearby
star image as the PSF: USNO-B1 0892-0038669 for \object{GRB~060904B}
and USNO-B1 0444-0107368 for \object{GRB~070420}. C magnitudes were
rescaled in the R band considering the absence of dramatic spectral changes
in the optical range.
In the case of \object{GRB~070420},
three stars lie very close to the GRB location:
USNO-B1 0444-0107222 (5", R=17.3),
USNO-B1 0444-0107218 (7", R=17.6),
USNO-B1 0444-0107203 (13", R=15.6).
We subtracted these stars using the PSF of
USNO-B1 0444-0107368 (R=13.3) before the PSF fitting
of the GRB.
Tables \ref{logobstable1} and \ref{logobstable2} give
photometric results of both GRBs from TAROT
and other useful measurements published in GCN Circulars.

\subsection{Trailed images}
\label{optprompt}

At the \object{GRB~060904B} position (Fig.~\ref{trail1} top),
no trail is detected at R$<$17.2
except for a small cluster
of 4 pixels near t$_\mathrm{trig}$+42.8\,sec. We first considered it
as a possible optical flash associated with the GRB
\citep{Klotz2006a}.
The punctual shape implies that its duration is less
than 3 seconds (the time sampling of the trail) corresponding to R$<$15.9.
Rykoff (private communication) provided us with a series of 5 second exposures
obtained with ROTSE-III telescopes. One of these images was taken between
t$_\mathrm{trig}$+42.6\,sec and t$_\mathrm{trig}$+47.6\,sec. It shows no object brighter
than R=16.7. As a consequence, we do not confirm this
flash to be real. From a statistical point of view, the signal 
to noise ratio (S/N) of the four involved pixels are : 1.0, 2.0, 2.7, 2.8. 
Considering a Gaussian noise distribution of the sky background, 
simultaneous values higher than the corresponding S/N is $\sim 10^{-6}$ 
for four contiguous pixels. This excludes a natural background fluctuation. 
Thus, it is probably an artifact ({\it e.g.} a cosmic trace).
We took this into account when we computed limiting magnitudes
(see Fig.~\ref{lc1}).
From TAROT and ROTSE-III data, we conclude that the optical emission was
R$>$17.2 during the prompt emission, except
during t$_\mathrm{trig}$+41.3\,sec to t$_\mathrm{trig}$+44.3\,sec
where it was R$>$15.9.
\\
In the case of \object{GRB~070420}
(Fig.~\ref{trail2} top), a very faint trace is present at the
beginning of the theoretical place of the trail but
the signal to noise ratio is lower than 3, preventing any
conclusion.
\\

%%%%%%%%%%%%%%%%%%%%%%%%%%% TABLE  %%%%%%%%%%%%%%%%%%%%%%%%%%%%%
\begin{table}[htb]
\caption{ Log of the optical measurements of \object{GRB~060904B} from TAROT and GCN
Circulars. T are
seconds since $t_{trig}$. C magnitudes of TAROT images were translated
to the R photometric band by an appropriate offset.
Errors $\Delta$mag are 2$\sigma$ for TAROT data.
}
\begin{center}
{\scriptsize
\begin{tabular}{c c c c c}
\hline\hline
T$_\mathrm{start}$ & T$_\mathrm{end}$ & mag. & $\Delta$mag. & GCNC ref. \cr
\noalign{\smallskip} \hline \noalign{\smallskip}
  18.5  &   23.5  & R 17.30   & 0.10    & 5504 \cr
  23.0  &   41.3  & C $>$17.2  &         & TAROT \cr
  41.3  &   44.3  & C $>$15.9  &         & TAROT \cr
  44.3  &   86.0  & C $>$17.2  &         & TAROT \cr
  57.8  &  147.8  & R 18.21   & 0.02    & 5511 \cr
  90.8  &  120.8  & C 18.38   & 0.47    & TAROT \cr
 128.1  &  157.4  & C 17.76   & 0.32    & TAROT \cr
 164.1  &  194.7  & C 17.99   & 0.35    & TAROT \cr
  71.0  &  316.0  & V 18.64   & 0.30    & 5519 \cr
 200.7  &  230.7  & C 17.61   & 0.26    & TAROT \cr
 237.8  &  267.9  & C 17.15   & 0.20    & TAROT \cr
 177.8  &  358.8  & B 18.32   & 0.04    & 5511 \cr
 283.5  &  373.5  & C 17.31   & 0.13    & TAROT \cr
 380.1  &  470.1  & C 17.02   & 0.11    & TAROT \cr
 387.8  &  477.8  & R 17.01   & 0.02    & 5511 \cr
 502.0  &  527.0  & R 16.90   & 0.12    & 5541 \cr
 479.7  &  569.7  & R 16.78   & 0.09    & TAROT \cr
 526.2  &  526.2  & R 16.75   & 0.02    & 5524 \cr
 597.6  &  597.6  & R 16.86   & 0.03    & 5524 \cr
 507.8  &  687.8  & B 17.51   & 0.04    & 5511 \cr
 578.0  &  668.0  & C 16.94   & 0.11    & TAROT \cr
 664.2  &  664.2  & R 16.99   & 0.03    & 5524 \cr
 653.0  &  678.0  & R 17.09   & 0.16    & 5541 \cr
 674.7  &  765.3  & C 17.06   & 0.11    & TAROT \cr
 735.0  &  735.0  & R 17.05   & 0.04    & 5524 \cr
 718.8  &  808.8  & R 17.11   & 0.02    & 5511 \cr
 784.8  &  784.8  & R 17.22   & 0.05    & 5524 \cr
 777.0  &  803.0  & R 17.24   & 0.19    & 5541 \cr
 774.3  &  864.3  & R 17.22   & 0.14    & TAROT \cr
 874.8  &  874.8  & R 17.34   & 0.06    & 5524 \cr
 889.0  &  914.0  & R 17.36   & 0.17    & 5541 \cr
 881.1  &  971.1  & C 17.52   & 0.17    & TAROT \cr
 838.8  & 1018.8  & B 18.56   & 0.04    & 5511 \cr
1014.0  & 1014.0  & R 17.48   & 0.09    & 5524 \cr
1004.0  & 1029.0  & R 17.44   & 0.14    & 5541 \cr
 978.3  & 1067.7  & C 17.54   & 0.16    & TAROT \cr
1083.0  & 1083.0  & R 17.50   & 0.09    & 5524 \cr
1073.0  & 1098.0  & R 17.58   & 0.14    & 5541 \cr
1048.8  & 1138.8  & R 17.79   & 0.02    & 5511 \cr
1077.3  & 1167.3  & R 17.34   & 0.16    & TAROT \cr
1153.8  & 1153.8  & R 17.64   & 0.10    & 5524 \cr
1175.6  & 1265.6  & C 17.56   & 0.16    & TAROT \cr
1168.8  & 1348.8  & B 18.81   & 0.04    & 5511 \cr
1272.8  & 1362.3  & C 17.93   & 0.22    & TAROT \cr
1371.3  & 1461.8  & R 17.86   & 0.24    & TAROT \cr
1378.8  & 1468.8  & R 18.21   & 0.02    & 5511 \cr
1466.0  & 1491.0  & R 17.70   & 0.22    & 5541 \cr
1505.0  & 1530.0  & R 17.75   & 0.20    & 5541 \cr
1479.3  & 1659.8  & C 17.63   & 0.17    & TAROT \cr
1669.0  & 1694.0  & R 17.94   & 0.18    & 5541 \cr
1665.8  & 1846.5  & C 17.96   & 0.20    & TAROT \cr
1712.8  & 1802.8  & R 17.66   & 0.02    & 5511 \cr
1751.0  & 1776.0  & R 17.79   & 0.21    & 5541 \cr
1802.0  & 1827.0  & R 17.89   & 0.14    & 5541 \cr
1854.8  & 2034.8  & R 17.67   & 0.17    & TAROT \cr
2044.5  & 2223.8  & C 17.78   & 0.20    & TAROT \cr
2231.1  & 2410.4  & C 17.96   & 0.18    & TAROT \cr
2242.8  & 2422.8  & R 17.78   & 0.02    & 5511 \cr
2419.5  & 2600.0  & R 17.61   & 0.22    & TAROT \cr
2538.8  & 2718.8  & R 17.79   & 0.02    & 5511 \cr
2805.8  & 2985.8  & R 17.99   & 0.02    & 5511 \cr
56592.0  & 56592.0  & R 21.40   & 0.20    & 5548 \cr
76529.4  & 77969.4  & R 21.63   & 0.18    & 5741 \cr
86477.4  & 87377.4  & R 21.75   & 0.11    & 5526 \cr
87514.2  & 88414.2  & R 21.74   & 0.12    & 5526 \cr
89415.0  & 90315.0  & R 21.80   & 0.20    & 5526 \cr
162531.0  & 163731.0  & R 22.40   & 0.30    & 5741 \cr
\noalign{\smallskip} \hline
\end{tabular}
}
\label{logobstable1}
\end{center}
\end{table}
%%%%%%%%%%%%%%%%%%%%%%%%%%%%%%%%%%%%%%%%%%%%%%%%%%%%%%%%

%%%%%%%%%%%%%%%%%%%%%%%%%%% TABLE  %%%%%%%%%%%%%%%%%%%%%%%%%%%%%
\begin{table}[htb]
\caption{ Log of the optical measurements of \object{GRB~070420} from TAROT and GCN
Circulars.
Columns and units are the same as Table~\ref{logobstable1}.
}
\begin{center}
{\scriptsize
\begin{tabular}{c c c c c}
\hline\hline
T$_\mathrm{start}$ & T$_\mathrm{end}$ & mag. & $\Delta$mag. & GCNC ref. \cr
\noalign{\smallskip} \hline \noalign{\smallskip}
  39.9  &   48.9  & C $>$ 16.75   &  & TAROT \cr
  48.9  &   57.3  & C $>$ 16.45   &  & TAROT \cr
  57.3  &   65.7  & C $>$ 16.25   &  & TAROT \cr
  65.7  &   74.7  & C $>$ 16.75   &  & TAROT \cr
  74.7  &   83.1  & C $>$ 16.75   &  & TAROT \cr
  83.1  &   91.5  & C $>$ 16.75   &  & TAROT \cr
  90.0  &  100.0  & V 17.90   & 0.10    & 6336 \cr
 101.1  &  131.0  & C 16.64   & 0.18    & TAROT \cr
 137.6  &  166.4  & C 16.46   & 0.13    & TAROT \cr
 109.0  &  208.0  & V 17.40   & 0.10    & 6336 \cr
 173.1  &  203.1  & C 16.03   & 0.11    & TAROT \cr
 296.1  &  386.1  & C 16.38   & 0.12    & TAROT \cr
 392.1  &  482.0  & C 16.73   & 0.17    & TAROT \cr
 491.6  &  582.3  & R 16.84   & 0.23    & TAROT \cr
 590.0  &  680.0  & C 17.05   & 0.31    & TAROT \cr
 693.0  &  703.0  & B 18.80   & 0.10    & 6336 \cr
 686.0  &  776.0  & C 17.05   & 0.26    & TAROT \cr
 890.0  &  980.1  & C 17.50   & 0.41    & TAROT \cr
 986.1  & 1075.5  & C 17.65   & 0.47    & TAROT \cr
10530.0  & 11154.0  & R 19.70   & 0.40    & 6334 \cr
12678.0  & 13308.0  & I 19.30   & 0.50    & 6334 \cr
\noalign{\smallskip} \hline
\end{tabular}
}
\label{logobstable2}
\end{center}
\end{table}
%%%%%%%%%%%%%%%%%%%%%%%%%%%%%%%%%%%%%%%%%%%%%%%%%%%%%%%%

%%%%%%%%%%%%%%%%%%%%%%%%%%%%%%% DISCUSSION %%%%%%%%%%%%%%%%%%%%%%%%%
\subsection{The optical afterglow of \object{GRB~060904B}}
\label{optearly06}

\begin{figure*}[htb]
\sidecaption
\includegraphics[width=12cm]{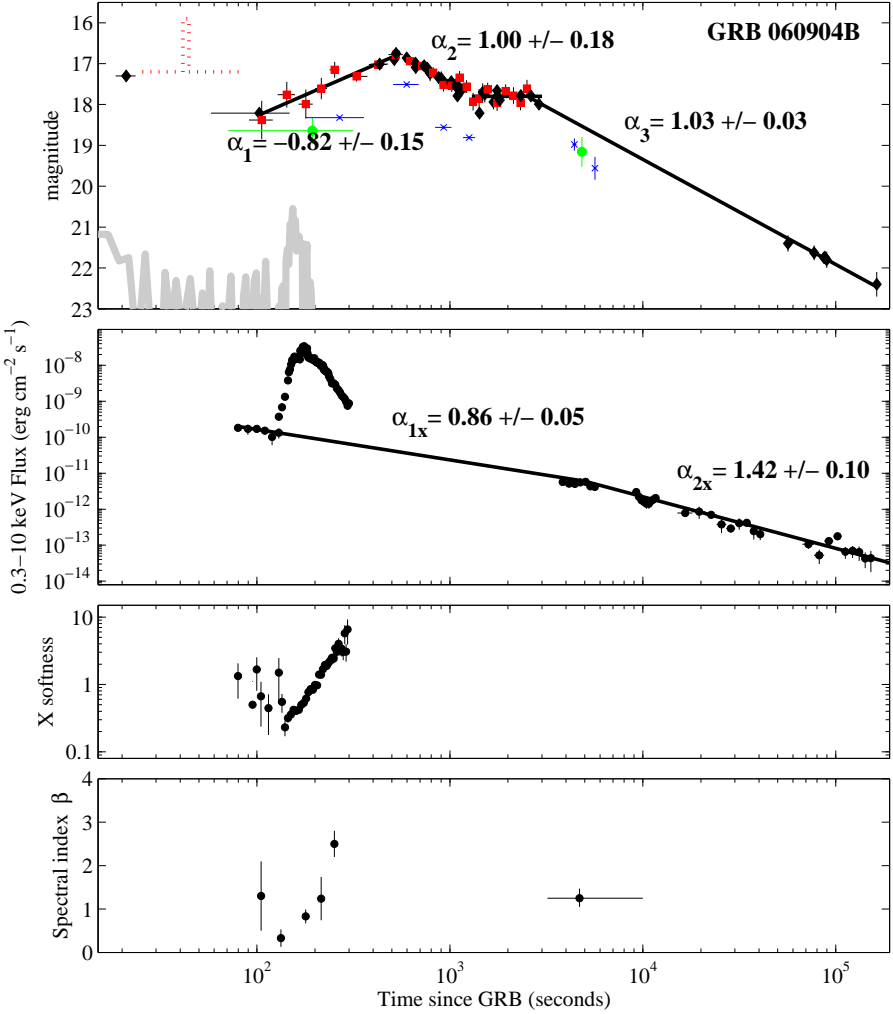}
\caption{
Time resolved parameters of \object{GRB~060904B}.
Top panel: Optical light-curve. Data are
provided from Table~\ref{logobstable1}.
Red squares are TAROT data (plus the limiting
magnitude during 23\,sec to 83\,sec indicated by the
dotted line).
Black diamonds are R measurements of other
observatories. Green disk and blue x are
from the literature for the V and B band respectively.
The BAT light curve is displayed as the light gray curve
offset arbitrarily.
Fits are based on the flux law: $f(t) \propto t ^{-\alpha}$.
Second panel: X-Ray light-curve from XRT data.
The count rate has been converted to flux units
using the best fit spectral model of late X-ray afterglow
(to avoid spurious features due to the large spectral
variations observed during the flare).
Third panel: softness ratio defined by (0.3-1.5 keV)/(1.5-10 keV).
Bottom panel: X-ray band spectral index.
} \label{lc1}
\end{figure*}

The optical light-curve of \object{GRB~060904B}
is shown in Fig.~\ref{lc1}. Before the end of the prompt phase,
the optical emission rose with a slope $\alpha_1$=-0.82
between t$_\mathrm{trig}$+90\,sec and t$_\mathrm{trig}$+550\,sec
reaching R=16.8 at t$_\mathrm{trig}$+550\,sec.
Then, the flux decreased
with a slope $\alpha_2$=1.0
down to R=17.8 at t$_\mathrm{trig}$ + 1270\,sec.
From this date, the flux remained at a constant level until
the end of the available early observations (t$_\mathrm{trig}$ + 49.8 min).
Late observations obtained 15\,hours after the trigger imply a late
decay index of $\alpha_3$=1.03 in the range t$_\mathrm{trig}$ + [50\,min - 1.9\,day].
\\
Interpolating R, B and V magnitude measurements,
we deduced (B-R) at several dates (Table~\ref{br}).
At $z$=0.703 (redshift of \object{GRB~060904B})
the color indexes are not affected by
the Lyman-$\alpha$ cut-off \citep{Roming2006}.
According to \citet{Simon2001}, (B-R)$_o$=0.8$\pm$0.3
for typical early afterglows.
The observed (B-R) values of \object{GRB~060904B},
corrected for the galactic extinction, are
compatible with a weak local extinction.
\citet{cobb07} obtained data in the JHK bands
during the late afterglow phase. They confirm
a weak extinction.

\subsection{The optical afterglow of \object{GRB~070420}}
\label{optearly07}

\begin{figure*}[htb]
\sidecaption
\includegraphics[width=12cm]{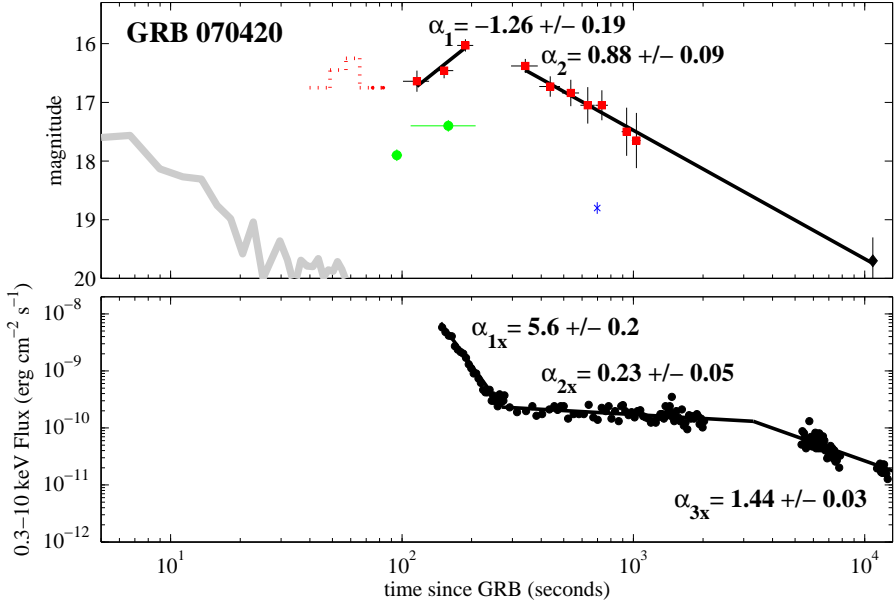}
\caption{
Global light-curves of \object{GRB~070420}.
Data are from Table~\ref{logobstable2}.
Red squares are TAROT data (plus the limiting
magnitude during 40\,sec to 91\,sec indicated by the
dotted line).
Black diamonds are R measurements of other
observatories. The green disk and blue x are
from the literature for the V and B band respectively.
The BAT light curve is displayed as the light gray curve.
} \label{lc2}
\end{figure*}

The optical light-curve of \object{GRB~070420}
is shown in Fig.~\ref{lc2}.
After the end of the prompt phase,
optical emission had risen at a rate $\alpha_1$=-1.26.
This rising rate is compatible with V band data provided
by UVOT \citep{Immler2007}.
The optical data show a temporal gap at the maximum
of the emission (due to a technical problem).
Extrapolating the data, we estimate the
peak of emission to occur near t$_\mathrm{trig}$+3.3min
with a magnitude of R=15.9.
Then the flux decreased
at a rate of $\alpha_2$=0.89
to R=17.6 at t$_\mathrm{trig}$+17.2\,min.
Later observations obtained 3\,hours after the trigger are compatible
with this decay.
If we set the t$_\mathrm{trig}$ time to the onset
of the start of the BAT emission (50\,sec earlier than the trigger time)
we obtain $\alpha_1$=-1.69 and $\alpha_2$=+0.91.
\\
Interpolating R and some other band magnitude measurements
we deduced color indexes at several epochs (Table~\ref{br2}).
At t$_\mathrm{trig}$+95\,sdec,
(V-R)=+0.5 (corrected for galactic extinction).
At t$_\mathrm{trig}$+698\,sec, during the optical
decreasing phase,
(B-R)=+0.85.

%%%%%%%%%%%%%%%%%%%%%%%%%%% TABLE  %%%%%%%%%%%%%%%%%%%%%%%%%%%%%
\begin{table}[htb]
\caption{Color indexes for \object{GRB~060904B}.
T$_\mathrm{mean}$ is the time
since t$_\mathrm{trig}$ when the color indice is measured.
T$_\mathrm{mean}$ is expressed in seconds.
Column 'mag.'
is not corrected for galactic extinction.
Column 'dereddened' is the color corrected by the
galactic extinction.
}
\begin{center}
{\scriptsize
\begin{tabular}{c c c c c c}
\hline\hline
T$_\mathrm{mean}$ & color & mag. & deredened & uncert. & remarks \cr
\noalign{\smallskip} \hline \noalign{\smallskip}
 193  & V-R  & 1.0    & 0.89 & 0.3     & bad sampling \cr
 268  & B-R  & 0.93   & 0.65 & 0.04    & optical rising \cr
 598  & B-R  & 0.72   & 0.44 & 0.04    & close to the top \cr
      &      &        &      &         & of the bump \cr
 929  & B-R  & 1.10   & 0.82 & 0.04    & optical decreasing \cr
1259  & B-R  & 1.0    & 0.72 & 0.04    & optical shallowing \cr
4425  & B-R  & 0.58   & 0.30 & 0.15    & late afterglow \cr
4831  & V-R  & 0.64   & 0.53 & 0.36    & late afterglow \cr
5649  & B-R  & 0.88   & 0.60 & 0.28    & late afterglow \cr
\noalign{\smallskip} \hline
\end{tabular}
}
\label{br}
\end{center}
\end{table}
%%%%%%%%%%%%%%%%%%%%%%%%%%%%%%%%%%%%%%%%%%%%%%%%%%%%%%%%

%%%%%%%%%%%%%%%%%%%%%%%%%%% TABLE  %%%%%%%%%%%%%%%%%%%%%%%%%%%%%
\begin{table}[htb]
\caption{
Color indexes for \object{GRB~070420}. See column descriptions in Table~\ref{br}.
}
\begin{center}
{\scriptsize
\begin{tabular}{c c c c c c}
\hline\hline
T$_\mathrm{mean}$ & color & mag. & deredened & uncert. & remarks \cr
\noalign{\smallskip} \hline \noalign{\smallskip}
 95   & V-R  & 0.8    & 0.5  & 0.2     & optical rising \cr
 300  & R-H  & 3.15   & 2.1  & 0.1     & optical decreasing \cr
 698  & B-R  & 1.7    & 0.85 & 0.1     & optical decreasing \cr
\noalign{\smallskip} \hline
\end{tabular}
}
\label{br2}
\end{center}
\end{table}
%%%%%%%%%%%%%%%%%%%%%%%%%%%%%%%%%%%%%%%%%%%%%%%%%%%%%%%%

%%%%%%%%%%%%%%%%%%% FIG light-curve %%%%%%%%%%%%%%%%%%%
%%%%%%%%%%%%%%%%%%%%%%%%%%%%%%%%%%%%%%%%%%%%%%%%%%%%%%%%

%%%%%%%%%%%%%%%%%%% FIG light-curve %%%%%%%%%%%%%%%%%%%
%%%%%%%%%%%%%%%%%%%%%%%%%%%%%%%%%%%%%%%%%%%%%%%%%%%%%%%%

\section{High energy observations}
\label{xearly}

\subsection{BAT analysis of \object{GRB~060904B}}
\label{bat06}

\begin{figure}[htb]
\centering{\includegraphics[width=0.7\columnwidth, angle=270]{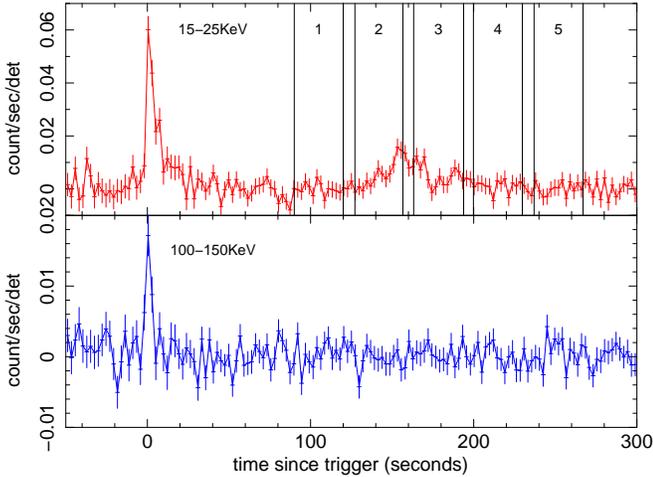}}
\caption{BAT light curves at different energy ranges for \object{GRB~060904B}.
The second peak near $t_{trig}$+160\,sec is only visible in the
softest energy band.
The vertical lines define the 5 intervals coincident with the TAROT observations.
}
\label{bat2}
\end{figure}

We have performed temporal and spectral analysis of the BAT data for
\object{GRB~060904B}. 
Figure~\ref{bat2} shows the light curve in the 15--25 and 100--150~keV ranges: the numbered vertical
columns refer to the TAROT observation periods with a detected optical source.
No high energy precursor is detectable up to t$_{trig}-200$s.
No signal was detected by BAT during the first TAROT observation (window number 1),
whereas observations number 2 and 3 cover the second and much fainter peak
in the high energy light curve.
In order to obtain broad band time-resolved spectra for \object{GRB~060904B}, we performed spectral
analysis of the BAT data in three different epochs: the main peak,
visible only by BAT (from t$_{trig}-5$s to t$_{trig}+25$s),
the rise of the second peak (from t$_{trig}+123.9$s to t$_{trig}+159.9$s), and the decay of the second peak
(from t$_{trig}+159.9$s to t$_{trig}+197.1$s). The second and third time intervals are slightly longer than (but centered on) the two corresponding TAROT observations (
t$_{trig} + [128.1 - 157.4]$s and t$_{trig} + [164.1 - 194.7]$s respectively).
This choice was made
in order to increase the signal to noise ratio and thus the quality of the obtained spectra while maintaining a good simultaneity of the observations.

The 15-150~keV energy spectra during all the 2 and 3 TAROT epochs are well fitted by a simple power law model.
We find clear evidence of spectral softening through the three epochs:
the photon index is in fact $\Gamma_1 = 1.35 \pm0.1$ during the main peak,
$\Gamma_2 = 2.16 \pm0.3$ during observation number 2 and $\Gamma_3 = 2.34 \pm0.4$ during observation number 3.
The spectral softening is also confirmed by a comparison of the light curves
in the 15-25~keV and 100-150~keV energy bands (Fig.~\ref{bat2}).

The fluence in the first peak is $1.22 \times 10^{-6}~$erg~cm$^{-2}$,
in the second peak (observations 2 and 3 together)
$3.68 \times 10^{-7}~$erg~cm$^{-2}$ for a total fluence of $1.6 \times 10^{-6}~$erg~cm$^{-2}$.

%%%%%%%%%%%%%%%%%%% FIG spectrum %%%%%%%%%%%%%%%%%%%
%%%%%%%%%%%%%%%%%%%%%%%%%%%%%%%%%%%%%%%%%%%%%%%%%%%%%%%%

\subsection{XRT analysis of \object{GRB~060904B}}
\label{xrt_060904b}

The XRT began observing 69 seconds after the BAT trigger \citep{Grupe2006}.
The data were processed using the ftools version 6.1.2. Data taken in {\it window timing} mode suffered from severe pile-up
during the flare and were corrected using the prescriptions indicated in \citet{Romano2006}.
We used a rectangle box extraction region excluding 2 central pixels for the
time intervals between
t$_\mathrm{trig}$+145\,sec and t$_\mathrm{trig}$+166\,sec and
between t$_\mathrm{trig}$+182.9\,sec and t$_\mathrm{trig}$+237.3\,sec,
and 5 central pixels for the time interval between
t$_\mathrm{trig}$+166.0\,sec and t$_\mathrm{trig}$+182.9\,sec.
The extraction region was 50 pixel large.
We corrected for the effects induced by a bad column and the pile-up excluded region located within the extraction region
by estimating from the Ancillary Response File the fraction of lost counts, and corrected
all count rates for this loss. The light curve was extracted 
within the 0.3-10.0 keV band and rebinned
in order to obtain at least 25 counts per bin. All 
decay indexes indicated below are derived by fiting power laws 
using the $\chi^2$ statistic. As the 0.3-2.0 keV range is not free 
from absorption, we prefered, for comparison with other works, 
to produce the figures using the standard 2.0-10.0 keV band (free of absorption).
We thus converted all 0.3-10.0 keV count rates into a standard 2.0-10.0 keV flux
by using the appropriate mean conversion factor derived from the spectral analysis
(see below).
The light curve shows an initial shallow
decay between 77.3~sec and $\sim$138~sec followed by a giant flare
(temporal index rise=-17.83$\pm$0.02, temporal decay index=6.1$\pm$0.4)
with a duration of 490\,sec (see Fig. \ref{lc1}). This flare has a 0.3-10.0 keV mean flux of $2.2\times10^{-9}$ erg
s$^{-1}$ cm$^{-2}$ and a total fluence of $8.8\times10^{-7}$ erg cm$^{-2}$.
The data feature a temporal gap between $\sim 300$ and $\sim 3500$ seconds. However,
the data taken before and after the flare are
consistent with a continuous and smooth power law decay with index
$\alpha_{1x}=0.86 \pm 0.05$ followed by a break near $(5 \pm 1) \times 10^3$~sec and
a final decay with index $\alpha_{2x}=1.42 \pm 0.1$.
We do not find evidence of a plateau phase in X-ray.

We extracted the XRT spectra during each TAROT temporal bin (see Table~\ref{table_spectre_x}),
plus a global spectrum in order to derive a mean count-to-flux conversion factor
for light curve conversions.
All spectra were rebinned in order to include 20 net counts within each bin,
and fit using the $\chi^2$ statistic. The
spectral model was composed of a power law continuum absorbed by our galaxy (Galactic hydrogen
column density fixed at $1.21\times10^{21}$ cm$^{-2}$, Dickey \& Lockman 1990) and
by the host galaxy at $z$=0.703. When possible, we fit the BAT and XRT spectra together.
During the flare, the spectrum softens but more interestingly the foreground
hydrogen column increases during the rising phase of the flare and then it decreases.
This is a quite uncommon behavior and we interpret it as an artificial effect
of a wrong assumption on the intrinsic spectral model. A Band model or cut-off power law provides a
more realistic behavior (see Table \ref{table_spectre_x}), with a constant extragalactic absorption
of $N_H=(8 \pm 2) \times 10^{21}$~H~cm$^{-2}$ (at $z$=0.703), and an hard-to-soft
behavior, characteristic of prompt related emission. Very similar results have been
obtained in the past for another giant X-ray flare
associated with \object{GRB~050502B} \citep{Falcone2006}. The difference with \object{GRB~060904B}
is that no rising optical afterglow was detected during the flare (or after).
The late XRT data of \object{GRB~060904B} (from $\sim 3700$ to $\sim 6000$ seconds)
are compatible with a simple absorbed power law (extragalactic
$N_H = (5 \pm 2) \times 10^{21}$~H~cm$^{-2}$,
$\beta = 1.3 \pm 0.3$, $\chi^2_\nu = 1.21$, 19 d.o.f.). Compared to the spectral
index found for the early XRT data (before the giant X-ray flare, see Table
\ref{table_spectre_x}), no significant changes are observed within the errors.

%%%%%%%%%%%%%%%%%%%%%%%%%%% TABLE  %%%%%%%%%%%%%%%%%%%%%%%%%%%%%
\begin{table}[htb]
\caption{
Spectral analysis of \object{GRB~060904B}. We indicate only the best fit models
(top panel, a cut-off power law; bottom panel, a Band model).
Column Bin refers to the TAROT measurements
defined in Section~\ref{bat06} and reported
in Figure~\ref{bat2}.
Bins 1, 4, and 5 are XRT data only, while bins 2 and 3 are XRT and BAT data fit
together.
Values between parentheses are set to the value indicated.
Extragalactic absorption is computed in the GRB rest frame.
The two spectral indexes of the Band function are given as low energy spectral index first, high energy spectral index after.
Several values are not constrained by the fit (the $E_0$ values for bins 1 and 5 with a cut-off power law).
All errors are given at the 90\% confidence level. When applicable, upper limits are given at the 95\% confidence level.}
\begin{center}
{\scriptsize
\begin{tabular}{c c c c c c c}
\hline\hline
Epoch               & Bin  & Spectral & $E_0$ & Extragalactic & $\chi^2_\nu$ & d.o.f. \cr
(seconds            &      & index    &  (keV)& absorption    &              &        \cr
since trigger)      &      &        &       &($10^{22}$ cm$^{-2}$)&        &        \cr
\noalign{\smallskip} \hline \noalign{\smallskip}
 $90.8-120.8$ & 1 &1.3 $\pm$ 0.8 & ---                    & (0.78)        & 0.92 & 1  \cr
$127.2-156.6$ & 2 & 0.3 $\pm$ 0.2 & $37.9^{+36.6}_{-14.5}$ & 0.8 $\pm$ 0.5 & 1.08 & 28 \cr
$163.2-193.8$ & 3 & 0.8 $\pm$ 0.2 & $13.2^{+7.0}_{-4.1}$   & 0.8 $\pm$ 0.2 & 1.18 & 119\cr
$199.8-229.8$ & 4 & 1.2 $\pm$ 0.5 & $6.8^{+13.4}_{-3.8}$   & 0.6 $\pm$ 0.2 & 0.93 & 61 \cr
$237.0-267.0$ & 5 & 2.5 $\pm$ 0.3 & ---                    & 0.8 $\pm$ 0.2 & 1.09 & 42 \cr
\noalign{\smallskip} \hline
$127.2-156.6$ & & 0.3 $\pm$ 0.2 & $39.6^{+35.7}_{-15.4}$ & 0.8 $\pm$ 0.5 & 1.12 & 27 \cr
              & & $> 6.84$      &                        &               &      &    \cr
$163.2-193.8$ & & 0.6 $\pm$ 0.4 & $5.7^{+4.2}_{-2.9}$    & 0.7 $\pm$ 0.2 & 1.05 & 118\cr
              & & 1.8 $\pm$ 0.3 &                        &               &      &    \cr
$199.8-229.8$ & & 1.3 $\pm$ 0.5 & $7.4^{+4.1}_{-3.9}$    & 0.6 $\pm$ 0.1 & 0.95 & 60 \cr
              & &    ---        &                        &               &      &    \cr
$237.0-267.0$ & &    ---        & $< 3.59$               & 0.7 $\pm$ 0.2 & 1.11 & 41 \cr
              & & 2.4 $\pm$ 0.2 &                        &               &      &    \cr
\noalign{\smallskip} \hline
\end{tabular}
}
\label{table_spectre_x}
\end{center}
\end{table}
%%%%%%%%%%%%%%%%%%%%%%%%%%%%%%%%%%%%%%%%%%%%%%%%%%%%%%%%

\subsection{BAT analysis of \object{GRB~070420}}
\label{bat_070420}

\begin{figure}[htb]
\centering{\includegraphics[width=0.7\columnwidth, angle=270]{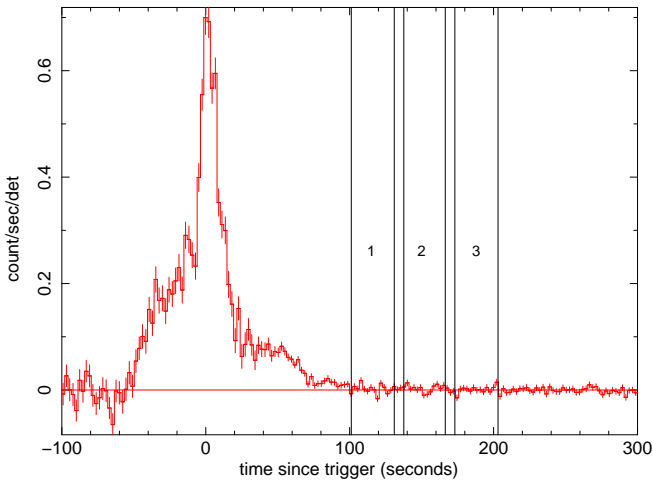}}
\caption{BAT total light curve (15-350 keV) for \object{GRB~070420}.
The vertical lines define the 3 intervals coincident with the TAROT observations.
Note that a signal due to \object{GRB~070420} is detected up to $\sim 50$ seconds before the trigger time.
}
\label{bat070420_2}
\end{figure}

We have performed temporal and spectral analysis of the BAT data for
\object{GRB~070420}.
The light curve, visible in Figure~\ref{bat070420_2} in the 15--350~keV energy range, shows a slow rise beginning about 50~sec before the BAT trigger,
then a multi-peaked structure and finally a gradual decay ending about 100~sec after the trigger. Again, the vertical columns refer to the TAROT observing windows, but for this GRB
no signal was detected by BAT that could be used to
derive simultaneous broad band spectra at those epochs.

We have therefore performed the spectral analysis
using the signal integrated over the BAT T$_{90}$: a simple power law fits the
observed spectrum between 15 and 150~keV with $\Gamma = 1.54 \pm0.06$,
while the burst fluence is $1.43 \times 10^{-5}$~erg~cm$^{-2}$.

We observed no significant spectral evolution during the emission detected by BAT.

%%%%%%%%%%%%%%%%%%% FIG spectrum %%%%%%%%%%%%%%%%%%%
%%%%%%%%%%%%%%%%%%%%%%%%%%%%%%%%%%%%%%%%%%%%%%%%%%%%%%%%

\subsection{XRT analysis of \object{GRB~070420}}
\label{xrt_070420}

The XRT observation starts 99.0 seconds after the BAT trigger;
and 159 seconds after the start of the
gamma-ray emission (which we define as the start of the event).
The 0.3-10.0 keV light curve is very smooth and features the standard ``steep-flat-steep'' behavior
observed in the \emph{Swift} era \citep{obr05}, with no flare superimposed
(see Fig. \ref{lc2}).
The three segments decay as $\alpha_{1x} = 5.6 \pm 0.2$,
$\alpha_{2x} = 0.23 \pm 0.05$
and $\alpha_{3x} = 1.44 \pm 0.03$ respectively, with two break times
of $t_{b1} = (3 \pm 1) \times 10^2$ and $t_{b2} = (3.3 \pm 0.2) \times 10^3$ seconds.
Note that these values differ from the ones reported in \citet{str07},
as the start time of the event is different \citep[][used the trigger time
as $T_0$, while we used the start of the event as $T_0$]{str07}.

The PC data suffer from a moderate pile-up during 280.2 and $\sim 7800$ seconds
after the trigger. The initial {\it window timing} data provide a good fit with
a simple absorbed power law, and are consistent with no spectral variations
(spectral index $\beta = 1.5 \pm 0.1$, extragalactic absorption
$N_H~<~1~\times~10^{20}$~H~cm$^{-2}$, $\chi_\nu^2 = 1.25$, 116 d.o.f.). The same spectral 
model results in a
spectral index of $\beta = 1.2 \pm 0.1$ and $\beta = 0.9 \pm 0.2$ ($\chi_\nu^2 = 1.30$, 50 d.o.f.)
during the plateau phase and the following decay respectively.
Note that the galactic hydrogen column density
along the direction of the GRB is $N_H = 3.7 \times 10^{21}~$H~cm$^{-2}$.

%%%%%%%%%%%%%%%%%%%%%%%%%%%%%
\section{Discussion}
\label{discussion}

\subsection{The giant flare of \object{GRB~060904B}}
\label{giant_flare}

Giant X-ray flares such as the one observed for \object{GRB~060904B} have ever
observed for other GRBs
\citep[see][for a review]{Burrows2007,Chincarini2007}.
In particular, the giant flare associated with \object{GRB~050502B} was studied in
detail by \citet{Falcone2006}.
Compared with the giant X-ray flare of \object{GRB~060904B}, the similarities are stricking.
\object{GRB~070704} is another example of a GRB with a large X-ray flare $\sim$300 seconds
after the trigger \citep{Godet2007}.

First, as for \object{GRB~050502B}, we find that the spectrum
of \object{GRB~060904B} during the flare cannot be fitted by a power law, unless
allowing for a variable absorption. The latter shows a significant
increase at the onset of the flare, followed by a decrease
(which is unusual behavior). We thus modeled the spectra using a
Band model and a cut-off model, finding a better explanation.
The consistency with the Band model (or cutoff power law)
suggests that the flare could be caused by the same mechanism powering
the prompt emission, as in a late time internal shock scenario,
which was suggested to explain the \object{GRB~050502B} giant flare.

The spectral and temporal analysis of the flare show other similarities
with the case of \object{GRB~050502B}, such as:
{\it i}) the increasing softening of
the spectrum as the flare develops (see Fig. \ref{lc1});
{\it ii)} the hard band decaying more quickly than the soft band
after the peak of the
flare ;
{\it iii)} the soft band having a shallower decay relative to the
slope of the flare rise (see Table \ref{table_decay_flare}).

The temporal decay observed before the flare seems to extrapolate
to $4000-5000$~sec after the trigger, with no evident increase in the
pre-flare versus the post-flare light curve normalization. This also suggests
that the X-ray flare could be produced by an additional mechanism, independent
to the one generating the underlying emission. Such emission, in turn,
could be interpreted as resulting from the superposition of a decaying prompt
and a rising X-ray afterglow component, as proposed by \cite{willingale07} to
interpret the typical ``steep-flat-steep'' behavior of \textit{Swift} XRT light
curves. In this scenario, the temporal break observed in the X-ray light curve
around $(5 \pm 1) \times 10^3$~sec could mark the transition from the ``flat phase''
to the ``steep'' afterglow-dominated phase. We note that the observed
steepening in the X-ray light curve ($\Delta\alpha = 0.56 \pm 0.11$) is not consistent with a spectral break
within a standard afterglow scenario (that would require $\alpha_{1x}-\alpha_{2x}=0.25$)
nor with an achromatic jet break (the optical light curve seems to extrapolate
without breaks from $\sim 3000$~sec to late times, see Fig. \ref{lc1}). In
addition, the late-time ($t\gtrsim 5000$~sec) optical-to-X-ray temporal and
spectral indices verify (within the errors) the closure relations expected
in a standard afterglow scenario, when $\nu_m$ is below the optical band
and $\nu_c$ between the optical and the X-ray band \citep[see e.g.][]{zha04}.

We thus conclude that the giant X-ray flare of \object{GRB~060904B} may be due
to a late internal shock.

%%%%%%%%%%%%%%%%%%% FIG spectrum %%%%%%%%%%%%%%%%%%%
\begin{figure}[htb]
\centering{\includegraphics[width=1.\columnwidth, angle=0]{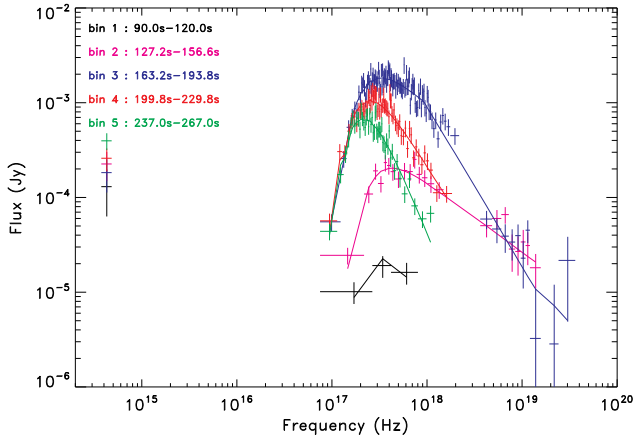}}
\caption{Evolution of the spectral emission distribution of \object{GRB~060904B} 
according to the
first bins corresponding to TAROT exposures. The theoretical model represented is a Band function.
It is not possible to reconcile optical and X-ray data
with only power-law and extinction.
This argues that the optical emission has a different 
origin to that of the X-ray flare.
}
\label{sed}
\end{figure}
%%%%%%%%%%%%%%%%%%%%%%%%%%%%%%%%%%%%%%%%%%%%%%%%%%%%%%%%

\begin{table}
\caption{
Summary of temporal indixes for the optical and X-ray light curves of \object{GRB 060904B}.
We report a decay with positive index and an increase with negative index.
}
\centering
\begin{tabular}{c c c}
\hline\hline
Part     & Optical index & X-ray index \cr
\noalign{\smallskip} \hline \noalign{\smallskip}
Pre X-flare & n.d. & $0.86 \pm 0.05$ \cr
X-flare rise & $-0.82 \pm 0.15$ & $-17.83 \pm 0.02$ \cr
X-flare decay & $-0.82 \pm 0.15$ & $6.1 \pm 0.4$ \cr
Optical Peak & 0 & $0.86 \pm 0.05$ \cr
First optical decay & $1.00 \pm 0.18$ & $0.86 \pm 0.05$ \cr
Optical plateau & 0 & $0.86 \pm 0.05$ \cr
Final decay before X-break & $1.03 \pm 0.03$ & $0.86 \pm 0.05$ \cr
Final decay after X-break & $1.03 \pm 0.03$ & $1.42 \pm 0.10$ \cr
\noalign{\smallskip} \hline
\end{tabular}
\label{table_decay_flare}
%\end{centering}
\end{table}

\subsection{Optical rising}
\label{11}

An optical rising has been observed for both \object{GRB~060904B} and \object{GRB~070420}.
For \object{GRB~070420}, the rising occured during the X-ray steep decay (see Fig.~\ref{lc2}).
V-band UVOT observations \citep{Immler2007} suggest an achromatic behavior.

Differently from the cases of \object{GRB~050502B} or \object{GRB~070704},
which had no optical counterpart,
for \object{GRB~060904B} an optical emission is observed:
the optical light curve rises smoothly during the whole X-ray flare,
and shows a maximum after the end of the X-ray flare.
One could consider the possibility that this flare
is correlated with the optical rise. In fact, the broad band
data suggest a temporal sequence, with a flare moving from the $\gamma$-ray
band to the optical band (see Fig.~\ref{lc1}). In such a case the X-ray and
optical emissions are produced by the same mechanism (internal shock, see
section~\ref{giant_flare}), with the peak of the emission moving from the BAT
to optical bands. Hence, near the peak of the emission one should expect
the same spectro-temporal behavior.
However, the observed X-ray rise and decay temporal slopes are significantly
steeper than the optical ones (see Table \ref{table_decay_flare}). This suggests that the optical rising and the
BAT-XRT flare are not correlated.
We also note that extrapolating in the optical band the best fit
Band models listed in Table \ref{table_spectre_x}, results in an overestimation
of the optical flux. We can reconcile the observed optical data with the high
energy extrapolation only by invoking a strong extinction. Moreover, such
extinction should increase as the flare develops, which is an unsual behavior.

A similar rise was observed in the near infrared light curves of \object{GRB~060418} and \object{GRB~060607A}
\citep{molinari2007}, together with a simultaneous X-ray emission characterized
by the presence of various flares. In these cases, the low-energy optical rise was
interpreted as the peak of the afterglow emission which, in the standard thin shell
case, is predicted to occur around the deceleration time $t_{dec}\sim t_{peak}$
\citep{Sari1999,Kobayashi2007}. Up to now, only a tenth of optical afterglows have been observed sufficiently early, and
time-sampled sufficiently well for the optical rising to be clearly identified.
Theoretically, if the rise has to be ascribed to the beginning of the afterglow, in a standard ISM scenario
the temporal index $\alpha_1$ is expected to be $\leq -2$ \citep{Sari1999},
much steeper than our observed value. In the case of a fireball expanding in a
wind environment, the rise could be less steep and also followed by a plateau
\citep[see e.g. Fig. 1 of ][]{Wu2003}. In such a case, we would expect the late optical
light curve to decrease with the same temporal index, or more steeply, than in the X-rays.
However, as discussed in section \ref{giant_flare}, if we interpret the X-ray emission
at $\sim 5000$~sec as the link to the ``standard'' afterglow phase, then the hypothesis
of an expansion in a wind environment does not agree with the late time observations.
In fact, at $t\gtrsim 5000$~sec, the X-ray decay is steeper than the optical one.

Other phenomena can also contribute to a late rising of the optical afterglow:
{\it i)} an off axis line of sight \citep{Granot2002}
as was supposed in the case of \object{GRB~060206} by
\citet{Wozniak2006} for which the fluence was very low
(8.4$\pm$0.4$\times$10$^{-7}$~erg~cm$^{-2}$ in the 15-150 keV range).
This is not the case for \object{GRB~070420} for which the fluence
is exceptionally high.
Moreover, in the case of an off-axis observer,
the X-ray and the optical light curves must follow
the same behavior, which is not the case of
the two studied GRBs.
{\it ii)} The circumstellar medium could be very dense and the
extinction very high at the beginning of the optical afterglow,
and could decrease later because of the dust destruction implied
by the burst blast wave \citep{Perna1998}.
This is probably not the case of \object{GRB~060904B} because
optical emission was detected by ROTSE during the
prompt emission and color indexes do not show
a red excess near $t_{peak}$.
Optical indexes prove that the extinction was not high in both GRBs.
{\it iii)} In the case of a smooth gradient of interstellar material density
\citep{Tam2005},
the rise is explained by an appropriate density profile
if the optical emission is below the cooling frequency
\citep{Sari1998}.
{\it  iv)} Reverse shock emission could be invoked.
In this scenario, one could interpret the optical bump
of \object{GRB~060904B}
as an optical flash, and the following plateau as the result of
the forward shock emission adding up to the reverse shock one:
see Fig. 2 of \citet{Sari1999}. However, in such a case, the
extrapolated power-law behavior of both the rise and decay of
the reverse shock optical emission are expected to be somewhat
steeper than those observed in our case \citep{Sari1999,Kobayashi2007}.
{\it  v)} the optical rise and decay
temporal indexes of $\alpha_1=-0.82\pm0.15$ and
$\alpha_2=1.00\pm0.18$, could be marginally accommodated
within a standard afterglow scenario, by having the $\nu_m$
cross the optical band.

\subsection{Optical plateau}
\label{plateau}

Some GRBs exhibit an optical plateau that consists of a phase
of very shallow decay starting a few minutes after the trigger
(typically 5~min in the burst rest frame) and lasting a few more minutes.
The plateau can appear simultaneously in the
optical and in the X-ray band \citep[{\it e.g} \object{GRB~050801}][]{depasquale07};
in some cases we
observe an early re-brightening rather than a plateau; in
other events a plateau appears in the optical on longer timescales, e.g
\object{GRB~060206} \citep{Stanek2007,monfardini06}
with no correlation with any X-ray flattening.

Of the two GRB analyzed in this work, only \object{GRB~060904B}
shows evidence of an optical plateau.
Several scenarios could explain such a plateau.
In the standard fireball scenario a nearly
flat optical light curve (temporal decay index $\alpha=~-0.25$) can
be produced by the forward shock when the fireball is in the fast
cooling regime with the observational frequency between the cooling
frequency $\nu_c$  and the injection frequency $\nu_m$, see Fig. 2 of
\citet{Sari1999}. For \object{GRB~060904B}, this ordering of characteristic
emission frequencies cannot be reconciled with the late ($t\gtrsim 5000$~sec) optical-to-X-ray data.
Another possibility is that the plateau is produced by a patchy jet, {\it i.e} by a collimated fireball
characterized by a non--uniform distribution of energy \citep{kumar00}.
In such a case we expect a large bump followed by a plateau
\citep[see Fig. 6 of][]{zhang06}, as we observe in \object{GRB~060904B}.
However, we should then see the same behavior roughly
simultaneously in all bands, which is
inconsistent with \object{GRB~060904B} X-ray data.
A last hypothesis for the origin of the optical plateau is
that it is produced by a late energy injection in the fireball.
Late energy injection can be ascribed to a long lasting central engine
activity, or to a refreshed shock associated with a short living
central engine that is releasing its energy with a variety of Lorentz
factors \citep{rees98}. This hypothesis was also proposed by
\citet{depasquale07} to explain the plateau observed simultaneously in
optical and X-ray in \object{GRB~050801}. In our case, it is difficult to conclude
about the presence of a simultaneous X-ray plateau, also because of the possible
contamination by internal shock emission.

We observe a large optical plateau when
a very large X-ray flare is observed (\object{GRB~060904B}) 
and no plateau when there
is no X-flare (\object{GRB~070420}). There are not enough observations
from other GRBs to confirm this point but this correlation
must be addressed by acquiring data of early afterglows
in both X-ray and optical wavelengths.

\section{Conclusion}

The optical emissions of
\object{GRB~060904B} and \object{GRB~070420}
are found to increase from the end of the prompt phase,
reaching a maximum of brightness at $t_{peak}$=9.2\,min. and 3.3\,min.
respectively and then decreasing.
\object{GRB~060904B} presents a large optical plateau and a huge X-ray flare.
We argue that the huge X-flare occurring near $t_{peak}$ is produced
by an extended internal engine activity. Its presence during the optical rise is only
a coincidence, and is not related to the optical flare. 
\object{GRB~070420} observations would support
this fact because there was no X-flare during the optical peak. We have proposed that the nature
of the optical plateau of \object{GRB~060904B},
while not completely elucidated, could be related to late energy injection.

\begin{acknowledgements}
B. Gendre acknowledges support from COFIN grant 2005025417.
The TAROT telescope has been funded by the {\it Centre National
de la Recherche Scientifique} (CNRS), {\it Institut National des
Sciences de l'Univers} (INSU) and the Carlsberg Fundation. It has
been built with the support of the {\it Division Technique} of
INSU. We thank the technical staff contributing to the TAROT project:
G. Buchholtz, J. Eysseric, M. Merzougui, C. Pollas, and P. \& Y. Richaud.
We thank E. Rykoff who provides some images of ROTSE-III.
S. Cutini, B. Preger \& G. Stratta acknowledge support
of ASI contract 1/024/05/0.

\end{acknowledgements}
%%%%%%%%%%%%%%%%%%%%%%%%%%%%%%%%% REFERENCES %%%%%%%%%%%%%%%%%%%%%%%

\end{document}